\documentclass[english,aps,prl,twocolumn,a4paper,superscriptaddress]{revtex4-1}
\usepackage[T1]{fontenc}
\usepackage[latin9]{inputenc}
\usepackage{geometry}
\usepackage{color}
\usepackage{babel}
\usepackage{textcomp}
\usepackage{amsmath}
\usepackage{amssymb}
\usepackage{graphicx}
\usepackage{esint}
\usepackage[unicode=true,pdfusetitle,
 bookmarks=true,bookmarksnumbered=false,bookmarksopen=false,
 breaklinks=false,pdfborder={0 0 1},backref=false,colorlinks=false]
 {hyperref}

\makeatletter

\newcommand{\noun}[1]{\textsc{#1}}

\makeatother

\begin{document}

\title{Search for light scalar Dark Matter candidate with AURIGA detector}

\author{Antonio Branca}

\affiliation{INFN, Sezione di Padova, Via Marzolo 8, I-35131 Padova, Italy;}

\author{Michele Bonaldi}

\affiliation{Institute of Materials for Electronics and Magnetism, Nanoscience-Trento-FBK
Division, I-38123 Trento, Italy; }

\affiliation{TIFPA - INFN, c/o Dipartimento di Fisica, Università di Trento, Via
Sommarive 14, 38123 Povo, Trento, Italy;}

\author{Massimo Cerdonio}

\affiliation{INFN, Sezione di Padova, Via Marzolo 8, I-35131 Padova, Italy;}

\author{Livia Conti}

\affiliation{INFN, Sezione di Padova, Via Marzolo 8, I-35131 Padova, Italy;}

\author{Paolo Falferi}

\affiliation{TIFPA - INFN, c/o Dipartimento di Fisica, Università di Trento, Via
Sommarive 14, 38123 Povo, Trento, Italy;}

\affiliation{Istituto di Fotonica e Nanotecnologie, CNR\textemdash Fondazione
Bruno Kessler, I-38123 Povo, Trento, Italy;}

\author{Francesco Marin}

\affiliation{Dipartimento di Fisica e Astronomia, Università di Firenze, Via Sansone
1, I-50019 Sesto Fiorentino (FI), Italy;}

\affiliation{Istituto Nazionale di Fisica Nucleare (INFN), Sezione di Firenze,
Via Sansone 1, I-50019 Sesto Fiorentino (FI), Italy; }

\affiliation{European Laboratory for Non-Linear Spectroscopy (LENS), Via Carrara
1, I-50019 Sesto Fiorentino (FI), Italy; }

\author{Renato Mezzena}

\affiliation{TIFPA - INFN, c/o Dipartimento di Fisica, Università di Trento, Via
Sommarive 14, 38123 Povo, Trento, Italy;}

\affiliation{Dipartimento di Fisica, Università di Trento, I-38123 Povo, Trento,
Italy;}

\author{Antonello Ortolan}

\affiliation{INFN - Laboratori Nazionali di Legnaro, I-35020 Legnaro (PD), Italy;}

\author{Giovanni A. Prodi}

\affiliation{TIFPA - INFN, c/o Dipartimento di Fisica, Università di Trento, Via
Sommarive 14, 38123 Povo, Trento, Italy;}

\affiliation{Dipartimento di Fisica, Università di Trento, I-38123 Povo, Trento,
Italy;}

\author{Luca Taffarello}

\affiliation{INFN, Sezione di Padova, Via Marzolo 8, I-35131 Padova, Italy;}

\author{Gabriele Vedovato}

\affiliation{INFN, Sezione di Padova, Via Marzolo 8, I-35131 Padova, Italy;}

\author{Andrea Vinante}

\affiliation{Istituto di Fotonica e Nanotecnologie, CNR\textemdash Fondazione
Bruno Kessler, I-38123 Povo, Trento, Italy;}

\author{Stefano Vitale}

\affiliation{TIFPA - INFN, c/o Dipartimento di Fisica, Università di Trento, Via
Sommarive 14, 38123 Povo, Trento, Italy;}

\affiliation{Dipartimento di Fisica, Università di Trento, I-38123 Povo, Trento,
Italy;}

\author{Jean-Pierre Zendri}

\affiliation{INFN, Sezione di Padova, Via Marzolo 8, I-35131 Padova, Italy;}
\begin{abstract}
A search for a new scalar field, called moduli, has been performed
using the cryogenic resonant-mass AURIGA detector. Predicted by string
theory, moduli may provide a significant contribution to the dark
matter (DM) component of our universe. If this is the case, the interaction
of ordinary matter with the local DM moduli, forming the Galaxy halo,
will cause an oscillation of solid bodies with a frequency corresponding
to the mass of moduli. In the sensitive band of AURIGA, some $100\,\mathrm{Hz}$
at around $1\,\mathrm{kHz}$, the expected signal, with a $Q=\tfrac{\triangle f}{f}\sim10^{6}$,
is a narrow peak, $\triangle f\sim1\,\mathrm{mHz}$. Here the detector
strain sensitivity is $h_{s}\sim2\times10^{-21}\,\mathrm{Hz^{-1/2}}$,
within a factor of $2$. These numbers translate to upper limits at
$95\%\,C.L.$ on the moduli coupling to ordinary matter $d_{e}\lesssim10^{-5}$
around masses $m_{\phi}=3.6\cdot10^{-12}\,\mathrm{eV}$, for the standard
DM halo model with $\rho_{DM}=0.3\,\mathrm{GeV/cm^{3}}$.
\end{abstract}
\maketitle
{\em Introduction} --- A possible source of Dark Matter (DM) is an
ultralight scalar field, $\Phi$, with couplings to Standard Model
(ordinary) matter weaker than the gravitational strength. For instance,
this field may be the moduli field, which is predicted by String Theory.
The coupling of this light field with ordinary matter implies a dependence
of the constants of nature on $\Phi$ \cite{DMSound}. In particular,
electron mass, $m_{e}$, and fine structure constant, $\alpha$, vary
with respect to their nominal values following:

\begin{equation}
m_{e}(\mathbf{x},t)=m_{e,0}\left[1+d_{m_{e}}\sqrt{4\pi G_{N}}\Phi(\mathbf{x},t)\right]\label{eq:me}
\end{equation}

\begin{equation}
\alpha(\mathbf{x},t)=\alpha_{0}\left[1+d_{e}\sqrt{4\pi G_{N}}\Phi(\mathbf{x},t)\right]\label{eq:alpha}
\end{equation}

where $G_{N}$ is the Newton's constant and $d_{m_{e}}$ ($d_{e}$)
is the dimensionless coupling of moduli to electrons (photons): $\Phi$
can be identified with an electron mass modulus if $d_{m_{e}}\neq0$
or an electromagnetic gauge modulus if $d_{e}\neq0$. Given relations
\ref{eq:me} and \ref{eq:alpha}, if the field $\Phi$ makes up a
significant fraction of the local DM density, the volume of a solid
will oscillate in time \cite{DMSound}. In fact, assuming the mass
of these particles, $m_{\Phi}$, to be small enough compared to the
energy density of the DM, their number density within our Galaxy is
high and the field $\Phi$ can be described as a classical wave, instead
of individual particles:

\begin{equation}
\Phi(\mathbf{x},t)=\Phi_{0}cos\left[m_{\Phi}(t-\mathbf{v\cdot x})\right]+O(v^{2})\label{eq:PhiClassic}
\end{equation}

where $\left|\mathbf{v}\right|$ is the relative velocity of DM with
respect to Earth, roughly equal to the virial velocity in our Galaxy.
Thus, the interaction of ordinary matter with the surrounding DM field
would make $m_{e}$ and $\alpha$ oscillate in time, causing a fluctuation
of the atoms size, $r_{0}\sim1/\alpha m_{e}$, in a solid. This would
imply a variation $\triangle L$ of the length of a body, corresponding
to a relative deformation with respect to its equilibrium length,
$L_{0}$, given by:
\begin{equation}
h(t)=\frac{\triangle L(t)}{L_{0}}=\frac{\sqrt{4\pi}}{M_{Pl}}(d_{m_{e}}+d_{e})\Phi(t)\label{eq:strain}
\end{equation}

where $M_{Pl}$ is the Plank mass and $\Phi(t)$ the moduli field.
To calculate the power spectrum of relative deformation $h$, we use
the so-called Standard Halo Model (SHM) that assumes a spherical DM
halo for the Galaxy with local DM density $\rho_{DM}=0.3\,\mathrm{GeV/cm^{3}}$,
and an isotropic Maxwell-Boltzmann speed distribution \cite{DMHaloMod}.
In this framework, if moduli account for a significant fraction of
DM in our Universe then the corresponding field $\Phi(t)$ can be
described as a zero mean stochastic process with a Maxwell-Boltzmann
power spectrum density \cite{ModuliModel}, consequently the spectrum
of the relative deformation $h$ is given by:
\begin{multline}
h(f)=\\
=h_{0}\frac{(d_{m_{e}}+d_{e})}{a^{\frac{3}{4}}f_{\phi}}(\left|f\right|-f_{\phi})^{\frac{1}{4}}e^{-\frac{(\left|f\right|-f_{\phi})}{2a}}\varTheta(\left|f\right|-f_{\phi})\label{eq:strainPWS}
\end{multline}

where $h_{0}=1.5\times10^{-16}\,\mathrm{Hz}$ is a constant, $f_{\phi}=m_{\Phi}/2\pi$
is the frequency corresponding to moduli with a given mass, $a=1/3f_{\phi}\left\langle v^{2}\right\rangle $
and $\left\langle v^{2}\right\rangle /c^{2}\sim10^{-6}$ the mean
squared velocity of the DM halo. Eq. \ref{eq:strainPWS} tells us
that the signal strain is a monopole (isotropic strain) and approximately
monochromatic.

\begin{figure}
\includegraphics[scale=0.29]{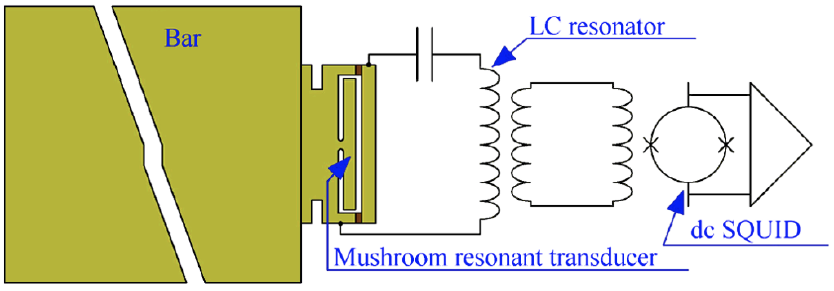}

\caption{(color online). Scheme of the gravitational wave detector AURIGA.
The system comprises three coupled resonators with nearly equal resonant
frequency of about $900\,\mathrm{Hz}$: the first longitudinal mode
of the cylindrical bar, the first flexural mode of the mushroom-shaped
resonator, which is also one of the plates of the electrostatic capacitive
transducer, and the low-loss electrical LC circuit. The electrical
current of the LC resonator is detected by a low noise dc SQUID amplifier.\label{fig:AURIGA}}
\end{figure}

\begin{figure}
\includegraphics[scale=0.2]{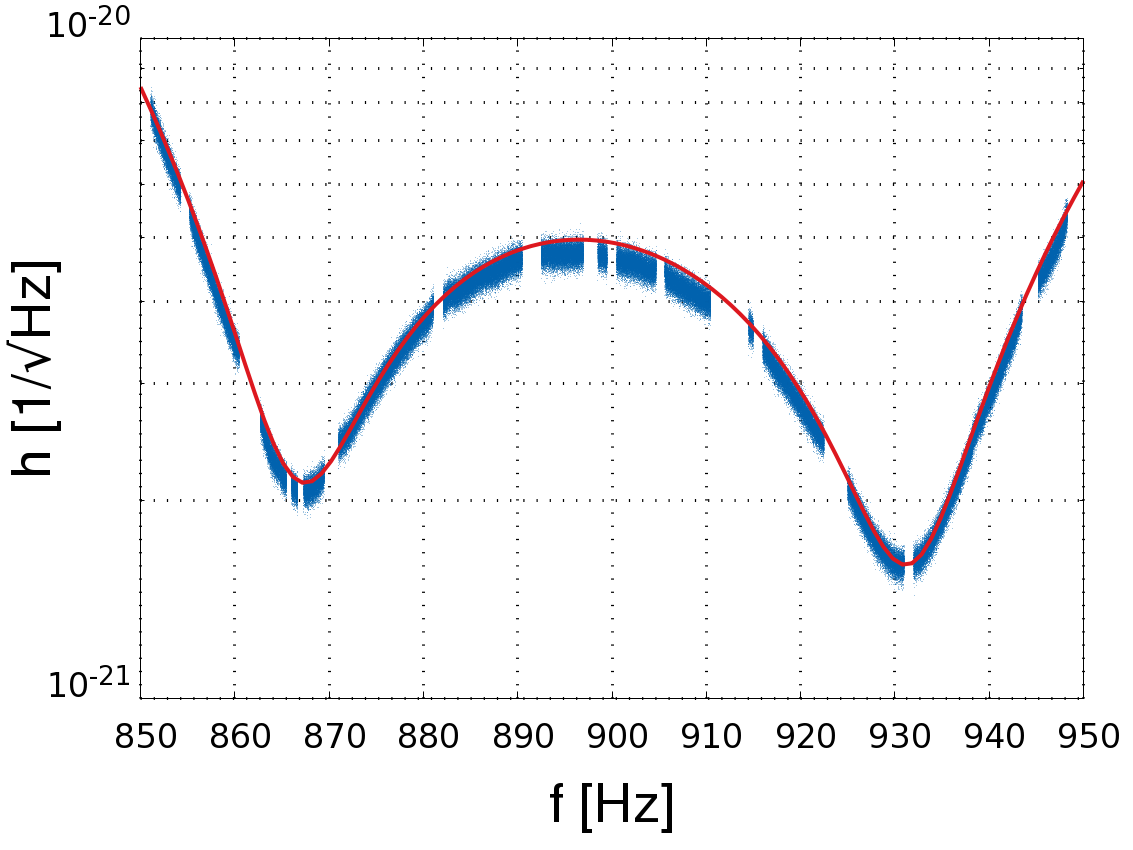}

\caption{(color online). Frequency spectrum of the bar relative deformation
computed on August 2015 AURIGA data (blue curve), obtained by averaging
$N=400$ power spectrums from $1$ hour long data streams. The experimental
result is compared to the predicted noise power spectrum density by
Fluctuation-Dissipation theorem (red line), showing a good matching.
The thickness of the data curve is due to the noise variance, reduced
by averaging the power spectrums. Holes in data correspond to excluded
spurious peaks associated to known external background. Moreover,
these spurious peaks have shape and width not matching the expectation
from moduli signal (see fig. (\ref{fig:Simulation})). \label{fig:strainDATA}}
\end{figure}

In this work, we analyze the data of the resonant-mass gravitational
wave detector AURIGA \cite{AURIGAdet}, searching for the strain induced
by an hypothetical moduli DM, expressed by eq. (\ref{eq:strainPWS}).
AURIGA represents the state-of-art in the class of gravitational wave
cryogenic resonant-mass detectors. It is located at INFN National
Laboratory of Legnaro (Italy) and has been in continuous operation
since year 2004. The detector is based on a $2.2\times10^{3}\,\mathrm{kg}$,
$3\,\mathrm{m}$ long bar made of low-loss aluminum alloy (Al5056),
cooled to liquid helium temperatures. The fundamental longitudinal
mode of the bar, sensitive to the moduli induced oscillation, has
an effective mass $M=1.1\times10^{3}\,\mathrm{kg}$ and a resonance
frequency $\omega_{B}/2\pi\simeq900\,\mathrm{Hz}$. The bar resonator
motion is detected by a displacement sensor with a sensitivity of
order several $10^{-20}\,\mathrm{m\cdot Hz^{-1/2}}$ over a $\thicksim100\,\mathrm{Hz}$
bandwidth. The spectral noise floor in the relative deformation for
the fundamental longitudinal mode, for the frequency interval of maximum
sensitivity is given in fig. (\ref{fig:strainDATA}). This sensitivity
is accomplished by a multimode resonant capacitive transducer \cite{Transd}
combined with a very low noise dc SQUID amplifier \cite{Squid} (Fig.
\ref{fig:AURIGA}). In this scheme, the bar resonator is coupled to
the fundamental flexural mode of a mushroom-shaped lighter resonator,
with $6\,\mathrm{kg}$ effective mass and the same resonance frequency.
As the mechanical energy is transferred from the bar to the lighter
resonator, the motion is magnified by a factor of roughly $15$. A
capacitive transducer, biased with a static electric field of $10^{7}\,\mathrm{V/m}$,
converts the differential motion between bar and mushroom resonator
into an electrical current, which is finally detected by a low noise
dc SQUID amplifier through a low-loss high-ratio superconducting transformer
\cite{LCresonator}. The transducer efficiency is further increased
by placing the resonance frequency of the electrical LC circuit close
to the mechanical resonance frequencies \cite{Transd}, at $930\,\mathrm{Hz}$.
The detector can then be simply modeled as a system of three coupled
resonators: its dynamics is described by three normal modes at separate
frequencies, each one being a superposition of the bar and transducer
mechanical resonators and the LC electrical resonator \cite{Modes}.

{\em Analysis workflow and data-set} --- Output from the read-out
chain of the AURIGA detector is digitized with a sampling frequency
of $f_{s}=4882.8\,\mathrm{Hz}$ through an ADC. As stated above, the
motion of the bar from the equilibrium length is converted into an
electrical signal. A calibration function obtained by a thorough mechanical
characterization of the system \cite{Modes} is then used to convert
data from electrical potential difference to relative deformation
$h$ of the AURIGA bar length. Eq. (\ref{eq:strainPWS}) shows that
the relative deformation induced by a signal moduli, would be a sharp
resonance around the frequency corresponding to the moduli mass. Therefore,
a possible signal could be spotted by analyzing the noise power spectrum
of the calibrated AURIGA output $P_{cal}(f)$. $P^{cal}(f)$ gives
the information concerning the relative deformation of the bar: 
\begin{equation}
h^{2}=\intop_{\triangle f}P^{cal}(f)df\label{eq:relStrainPWS}
\end{equation}
The expected signal (\ref{eq:strainPWS}) has a bandwidth of about
$\triangle f\sim1\,\mathrm{mHz}$ in the sensitive band of AURIGA.
Therefore, we split the analyzed dataset into one hour long data streams
and perform power spectrum computation on each stream to achieve the
proper spectrum resolution. Computed power spectrums are averaged
to reduce the noise standard deviation and achieve a better sensitivity.
If $N$ is the number of averaged power spectrums, the variance of
the noise is $N^{1/2}$ \cite{ProakisManolakis}, and the corresponding
standard deviation on $h$ decreases with the number of averages as
$N^{1/4}$. Thus, a good sensitivity on the moduli signal is already
achieved with few weeks of data. Using the entire dataset acquired
by AURIGA ($\sim10$ years) would improve the sensitivity just by
a factor of 3. So that, for this analysis we focused on a dataset
corresponding to data acquired during August 2015. AURIGA detector
has been running in stable conditions during this period: stability
of the detector is inferred by the stable frequencies and shape of
the three main detector's modes, checked by studying the evolution
of the detector power spectrum on the analyzed dataset. Spikes in
time due to energetic background events could hide a possible signal
from moduli and must be removed from the dataset: for each data stream
in the time domain the rms is computed, obtaining a distribution of
the rms value for the whole dataset; data affected by energetic background
lie in the high value tail of the distribution. A cut on the rms is
then set to discard data with large rms values. This cut still allows
to maintain a $86\%$ duty-cicle of the detector.

After cleaning data streams with rms cut, they are windowed in time
domain using a Hann window type, which allows a good frequency resolution
and reduced spectral leakage. The measured bar relative deformation
spectrum is shown in fig. (\ref{fig:strainDATA}). 
As shown by the figure, the measured noise is in excellent agreement
with the predicted noise behaviour. The latter has been obtained out
of the sum of computed contributions from each noise source, in turn
derived by measured experimental parameters \cite{Modes}. Few spurious
peaks, known to be associated to external background sources, have
been excluded from the analysis. 

\begin{figure}
\includegraphics[scale=0.21]{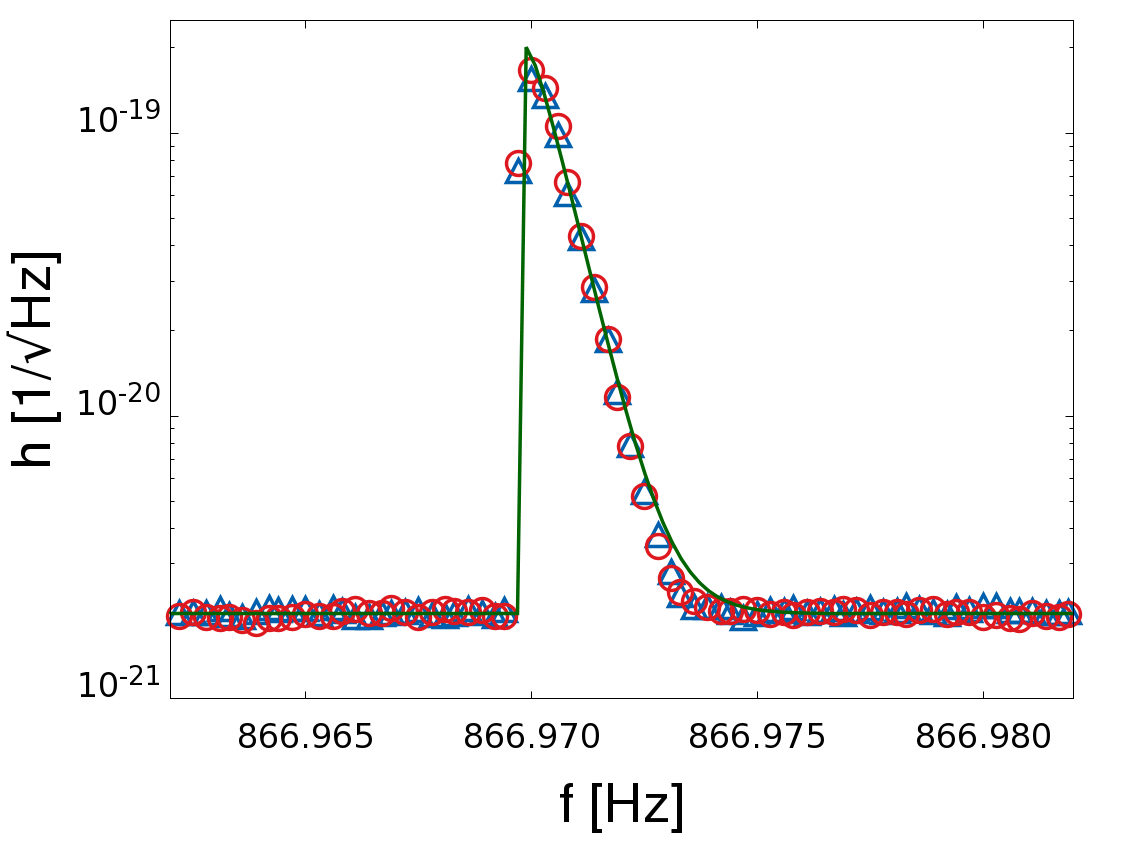}

\caption{(color online). (blue-triangle) Simulation of a moduli signal with
coupling $d_{m_{e}}=5\cdot10^{-4}$ and frequency $f_{\phi}\backsimeq867\,\mathrm{Hz}$
plus white noise with standard deviation $\sigma=2\cdot10^{-21}\,\mathrm{Hz^{-1/2}}$,
equal to the detector noise level at $f_{\phi}$. (red-circle) Same
simulated signal injected into the real data. The signal is a narrow
peak with a $\triangle f\backsimeq1\,\mathrm{mHz}$ bandwidth and
spread around about $10$ bins. (green-line) Plot of the power density
spectrum in eq. \ref{eq:strainPWS} plus a constant accounting for
the white noise with same parameters of the simulation.\label{fig:Simulation}}

\end{figure}

{\em Simulation} --- To prove we are able to detect this signal with
AURIGA, a simulation has been performed to study the actual signal
bandwidth within the detector sensitive region and to fine-tune the
analysis workflow. Eq. (\ref{eq:strainPWS}) is exploited to simulate
a signal with $f_{\phi}\backsimeq867\,\mathrm{Hz}$ and coupling $d_{m_{e}}=5\cdot10^{-4}$,
which is smaller than the natural values expected for $d_{m_{e}}$
\cite{DMSound}. $f_{\phi}$ lies close to the first minimum of the
AURIGA noise curve, shown in fig. (\ref{fig:strainDATA}). Given the
narrow bandwidth of this signal, we assumed the noise to be white,
$\left\langle n_{i}\right\rangle =0,\:\left\langle n_{i}n_{j}\right\rangle =\sigma^{2}\delta_{ij}$,
around the signal peak, with a standard deviation $\sigma=2\cdot10^{-21}\,\mathrm{Hz^{-1/2}}$,
equal to the noise level at $f_{\phi}\backsimeq867\,\mathrm{Hz}$
(see fig. \ref{fig:strainDATA}). We have generated an amount of data
comparable to the real dataset and applied our analysis pipeline obtaining
the result shown in fig. (\ref{fig:strainDATA}). The spectrum of
the simulated signal is spread around $\sim10$ bins of the spectrum
as shown in fig. (\ref{fig:Simulation} - \textit{\textcolor{black}{blue-triangles}}).
The simulated data have been injected into the real dataset and in
fig. (\ref{fig:Simulation} - \textit{\textcolor{black}{red-circles}})
we show that the injected signal is well reconstructed at the frequency
$f_{\phi}$ and it is not removed by the rms cut applied to the data
streams. We also show the theoretical signal plus noise, fig. (\ref{fig:Simulation}
- \textit{\textcolor{black}{green-line}}), obtained using same parameters
as for the simulation. The little discrepancy between theory and simulation
(injection), can be attributed to the minimal leakage due to the windowing
of data. 

\begin{figure}
\includegraphics[scale=0.16]{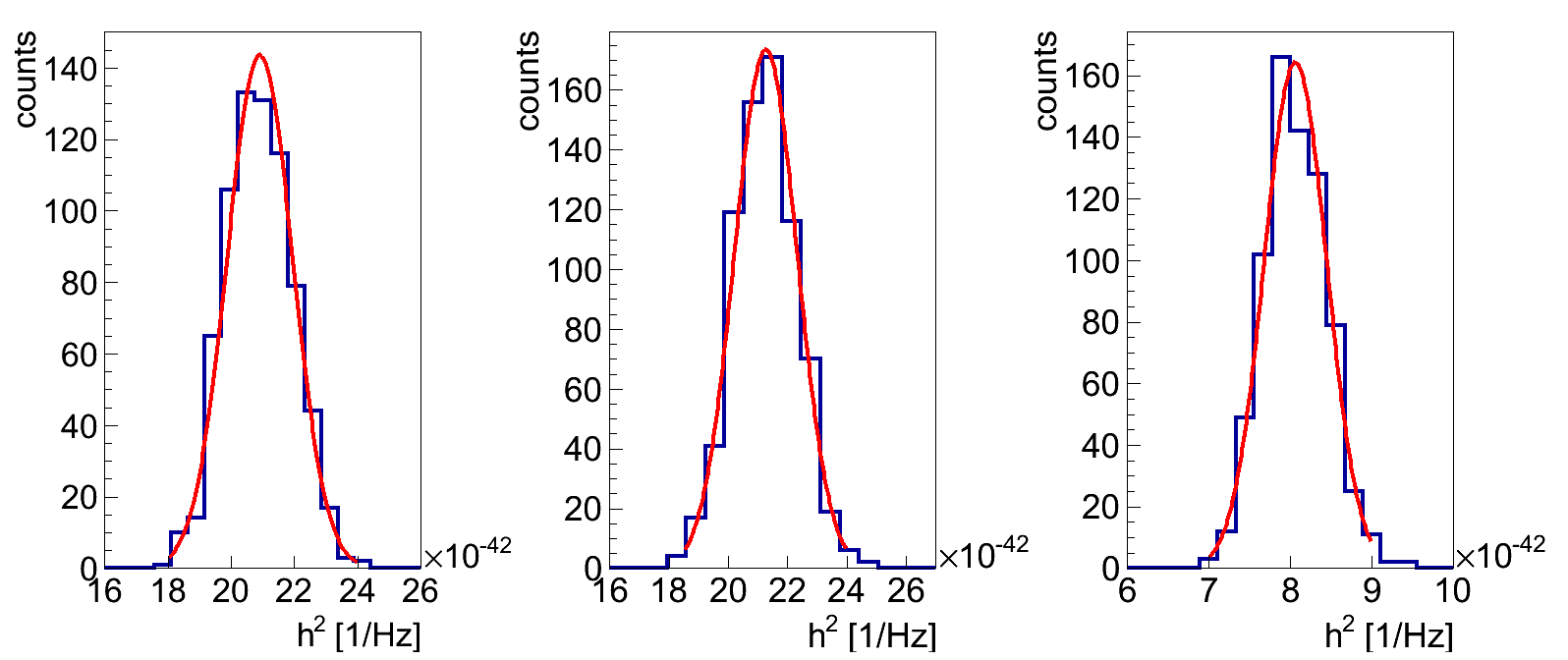}

\caption{(color online). Distributions of the possible values of three generic
bins in the relative bar deformation spectrum shown in fig. (\ref{fig:strainDATA}).
The distribution are well fitted by a gaussian (red lines) with mean
equal to the noise level at the considered bin frequency and standard
deviation the standard deviation of the noise: (left) $f=857\,\mathrm{Hz}$,
$\left\langle h^{2}\right\rangle =2.1\times10^{-41}\,\mathrm{Hz^{-1}}$,
$\sigma_{h^{2}}=1.0\times10^{-42}\,\mathrm{Hz^{-1}}$, $\chi^{2}/ndf=12.2/8$;
(center) $f=890\,Hz$, $\left\langle h^{2}\right\rangle =2.1\times10^{-41}\,\mathrm{Hz^{-1}}$,
$\sigma_{h^{2}}=1.1\times10^{-42}\,\mathrm{Hz^{-1}}$, $\chi^{2}/ndf=4.8/5$;
(right) $f=940\,\mathrm{Hz}$ $\left\langle h^{2}\right\rangle =8.1\times10^{-42}\,\mathrm{Hz^{-1}}$,
$\sigma_{h^{2}}=3.8\times10^{-43}\,\mathrm{Hz^{-1}}$, $\chi^{2}/ndf=8.3/7$.\label{fig:binStat}}
\end{figure}

{\em Statistical analysis} --- The procedure followed for the statistical
analysis of the result shown in fig. (\ref{fig:strainDATA}) is the
one proposed by Feldman and Cousin \cite{upperlFC}. Each bin of the
distribution in fig. (\ref{fig:strainDATA}) has a contribution from
the noise and a possible contribution from the signal. The squared
value of a bin is the result of averaging $N$ power spectrums, then
its distribution follows a non-central $\chi^{2}$ with $N$ degree
of freedom. Since in our case $N\sim400$ the squared bin distribution
can be approximated by the following gaussian: 
\begin{equation}
P(\overline{x}|\mu)=C\cdot exp\left[-\frac{(\bar{x}-\sigma^{2}-\mu^{2})^{2}}{\frac{2}{N}\sigma^{4}\left(1+2\frac{\mu^{2}}{\sigma^{2}}\right)}\right]\label{eq:binStat}
\end{equation}
with normalization factor:
\begin{equation}
C=\frac{N^{1/2}}{\sigma^{2}\sqrt{2\pi\left(1+2\frac{\mu^{2}}{\sigma^{2}}\right)}}\label{eq:PDFNorm}
\end{equation}

where $\bar{x}$ is the squared bin content, $\sigma^{2}$ is the
expected noise level and $\mu$ the signal strength. The statistical
behavior of the bins in distribution of fig. (\ref{fig:strainDATA})
is confirmed by data as predicted by eq. (\ref{eq:binStat}). This
is shown in fig. (\ref{fig:binStat}). By means of eq. (\ref{eq:binStat})
we build the confidence belt in the parameter space $(\bar{x},\mu^{2})$,
delimited by the values $(x_{1}(\mu),x_{2}(\mu))$ such that:
\begin{equation}
\int_{x_{1}(\mu)}^{x_{2}(\mu)}P(\bar{x}|\mu)d\bar{x}=\alpha\label{eq:binStat3}
\end{equation}

for each value of the signal strength $\mu$ and a confidence level
$\alpha=0.95$. The contributions to the integral in eq. (\ref{eq:binStat3})
are ordered following a specific ordering function, as reported in
\cite{upperlFC}, in order to avoid problems on the parameter estimation
near the physical bounds of such parameters. Eq. (\ref{eq:binStat3})
states that for a fixed hypothetical signal strength $\mu$, the observed
value of the bin content $\bar{x}$ falls within the interval $(x_{1}(\mu),x_{2}(\mu))$
with a probability equal to $\alpha$. Thus, for each measured value
of $\bar{x}$ the upper and lower limits on the measured signal strength,
containing the true value $\mu$ with a $95\%$ probability, is obtained
by inversion of the constructed confidence belt. We set a threshold,
$\bar{x}_{th}$, corresponding to a maximum false alarm probability
of finding a signal, which is not actually there, equal to $3$ standard
deviations away from the background only hypothesis. For observed
values of $\bar{x}$ below $\bar{x}_{th}$ we set an upper limit on
the signal strain. Values above the threshold $\bar{x}_{th}$ would
correspond to an observed signal. Since in our measurement in fig.
(\ref{fig:strainDATA}) we do not observe values exceeding the threshold,
we set upper limits on $h$ at $95\%$ confidence level. Interpreting
these upper limits as given by moduli through eq. (\ref{eq:strainPWS}),
we convert these values in upper limits on the moduli coupling $d_{m_{e}}$
to ordinary matter. To improve the upper limits, we exploited the
noise curve obtained adding the thermal noise prediction from Fluctuation-Dissipation
theorem and the noise contribution from the SQUID. By performing a
least squares fit of data in fig. (\ref{fig:strainDATA}), we obtained
the upper limits at $95\%\,C.L.$ from the $\chi^{2}$ distribution.
This allows to get better upper limits by taking into account a more
precise estimation of errors from the fit. Further improvement is
obtained by averaging bins in groups of $10$ for data in fig. (\ref{fig:strainDATA}),
since the signal would be distributed around $\sim10$ bins, as shown
by fig. (\ref{fig:Simulation}). 

\begin{figure}
\includegraphics[width=1\columnwidth]{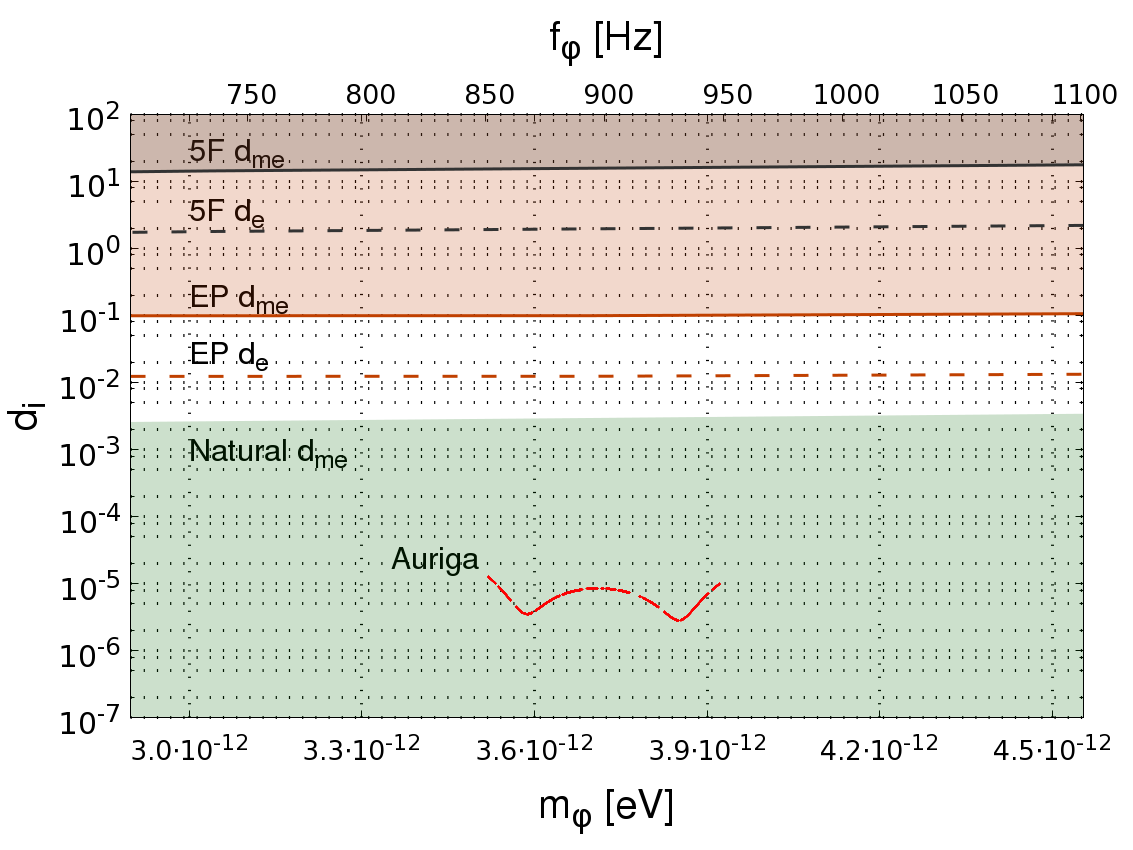}

\caption{Upper limits on the coupling of both an electron mass modulus ($d_{i}=d_{m_{e}}$)
and an electromagnetic gauge modulus ($d_{i}=d_{e}$) to ordinary
matter (red-curve) obtained from AURIGA data and reported in the moduli
parameter space: bottom and top horizontal axes represent the moduli
mass $m_{\phi}$ and corresponding frequency $f_{\phi}=m_{\phi}/2\pi$,
vertical axis represents the moduli coupling $d_{m_{e}}$ values.
Depicted green area shows the natural parameter space preferred by
theory. Other regions and dashed curves represent $95\%\,C.L.$ limits
on fifth-force tests (5F, gray) and equivalence-principle tests (EP,
orange). \label{fig:upperLimits}}
\end{figure}

{\em Results} --- Final upper limits are reported in fig. (\ref{fig:upperLimits}).
The upper limits set on the moduli coupling to ordinary matter are
better then $d_{i}\simeq10^{-5}$ in the sensitive band of AURIGA,
$\triangle f=[850,950]\,\mathrm{Hz}$, and explore an interesting
physical region of the parameter space, within the natural parameter
space for moduli \cite{DMSound}. With this result we prove that AURIGA,
a gravitational wave resonant detector, would be capable to detect
light DM candidates with an interesting sensitivity within its bandwidth.
We point out that this level of sensitivity can be achieved only by
resonant mass detectors, and not by modern laser interferometers developed
for gravitational wave detection, such as LIGO \cite{LIGO} and Virgo
\cite{VIRGO}, even if these have better sensitivity than resonant
mass detectors for gravitational waves and recently observed the first
event due to a gravitational wave signal \cite{GWDiscovery}. In fact,
because of the monopole nature of the expected moduli strain, we do
not expect an interference signal as output from the interferometer
due to moduli. Instead, since ultralight scalars can mediate Yukawa
forces between objects, one can explore which is the expected effect
on the relative position between mirrors within an interferometer
arm \cite{DMClocks}. The moduli signal could be measured as a difference
in the travel time between the mirrors. It turns out that the sensitivity
is masked by detector noise, therefore resulting in a lower searching
power with respect to resonant mass detectors.

\medskip{}

\begin{acknowledgments}
MC is very grateful to Asimina Arvanitaki for calling attention to
the matter and for initial discussions. AB, MC, AO, and LT thank Asimina
Arvanitaki and Ken Van Tilburg for enlightening discussions and for
a critical reading of the manuscript.\end{acknowledgments}

\end{document}